\documentclass[twocolumn,prl,aps,superscriptaddress,showpacs,amsmath,amssymb,floatfix]{revtex4-1}
\usepackage{graphicx}
\usepackage{amssymb}
\usepackage{amsmath}
\usepackage{epsfig}
\usepackage{color}
\usepackage{tabu}
\usepackage{mathtools}
\usepackage[colorlinks,linkcolor=blue,anchorcolor=blue,citecolor=blue,urlcolor=blue]{hyperref}
\usepackage{physics}
\usepackage{float}
\usepackage{diagbox}
\DeclareMathOperator{\Hc}{H.c.}

\begin{document}

\title{Cavity-mediated oscillating and trapping dynamics in a two-component condensate}
\author{Yue-Xin Huang}
\affiliation{Key Lab of Quantum Information, Chinese Academy of Sciences, School of physics, University of Science and Technology of China, Hefei, 230026, P.R. China}
\author{Zhen Zheng}
\affiliation{Key Lab of Quantum Information, Chinese Academy of Sciences, School of physics, University of Science and Technology of China, Hefei, 230026, P.R. China}
\author{Wei Feng Zhuang}
\affiliation{Key Lab of Quantum Information, Chinese Academy of Sciences, School of physics, University of Science and Technology of China, Hefei, 230026, P.R. China}
\author{Guang-Can Guo}
\affiliation{Key Lab of Quantum Information, Chinese Academy of Sciences, School of physics, University of Science and Technology of China, Hefei, 230026, P.R. China}
\affiliation{Synergetic Innovation Center of Quantum Information and Quantum Physics, University of Science and Technology of China, Hefei, 230026, P.R. China}
\author{Ming Gong}
\email{gongm@ustc.edu.cn}
\affiliation{Key Lab of Quantum Information, Chinese Academy of Sciences, School of physics, University of Science and Technology of China, Hefei, 230026, P.R. China}
\affiliation{Synergetic Innovation Center of Quantum Information and Quantum Physics, University of Science and Technology of China, Hefei, 230026, P.R. China}
\date{\today }

\begin{abstract}
Cold atoms in cavity provides a new platform for exploring exotic many-body phases. Here we explore the dynamics
of a two-component condensate coupled to a finesse cavity, in which the Raman coupling is mediated by pumping laser and cavity mode.
In this model, the energy scale of cavity mode is several order of magnitude bigger than that in the condensate,
thus the small fluctuations in the cavity field may have important consequence in the dynamics of condensate.
Beyond the steady-state approximation,
we show the cavity can play two different roles to this dynamics. In the first case, it imprints a gauge
potential to the dynamics of condensate, giving rise to zero and $\pi$ Josephson dynamics. Nevertheless,
in the other case, it plays the role of non-reciprocial transportation between the two hyperfine states,
in which the stability of the fixed points are tuned from elliptic to stable
spiral for one of the trapped phase and unstable spiral for the other trapped phase, thus the oscillating
dynamics will finally ceased. The transition between these dynamics can be controlled by the parameters of
the cavity field and the driving field.  Our results demonstrate an novel way to engineer the dynamics of  
condensate by tuning the stability of the fixed points.
\end{abstract}

\maketitle

Cold atoms trapped in a finesse cavity provides an ideal platform for exploring exotic many-body physics and associated dynamics
\cite{ritsch_cold_2013, kippenberg_cavity_2008}. In a finesse cavity, the coupling between cavity and the condensate
\cite{davis_bose-einstein_1995,anderson_observation_1995} can be much larger than the dephasing rate,
giving rise to strong coupling \cite{brennecke_cavity_2007,brennecke_cavity_2008,horak_coherent_2000}.
For a quantized cavity, this coupling can lead to superradiant phase \cite{kaluzny_observation_1983,mlynek_observation_2014,colombe_strong_2007}
and supersolid phase \cite{gopalakrishnan_atom-light_2010,li_lattice-supersolid_2013,leonard_supersolid_2017}. When an optical
lattice is coupled to a cavity, it is possible to realize the collective Dicke model, which has been long-sought in particle physics
and quantum optics for more than half-century. Recently, this model was realized by \citet{baumann_dicke_2010},
characterized by superradiant leaking of photon from the cavity. In the insulator regime with cavity-mediated long-range interaction, this
platform can host supersolid phase and Mott insulator phase. More intriguing phenomena such as strong nonlinear induced
bistability \cite{kartashov_bistable_2017,kazemi_controllable_2016,zhang_nonlinear_2009,zhou_cavity-mediated_2009,dong_multistability_2011,purdy_tunable_2010},
superradiant induced gauge potential \cite{jaksch_creation_2003,wang_superradiance_2015}, topological Kondo phase \cite{zheng_synthetic_2018}
and even topological superradiant phase \cite{nataf_no-go_2010} can also be realized. The similar physics may also be
explored in Fermi gases trapped in a cavity \cite{chen_superradiance_2014,pan_topological_2015,dong_photon-induced_2015,
dong_dynamical_2015,dong_cavity-assisted_2014,zhou_dynamically_2017,chen_superradiant_2015}.

In this work, we investigate the dynamics of a two-component condensate in a finesse cavity \cite{byrnes_neural_2013,law_stable_2010,
ye_dressed_2018, zhao_collective_2017,lode_many-body_2018}, in which the Raman coupling between the two hyperfine states is mediated 
by pumping laser and cavity mode. We find that the energy scale of the cavity field is much bigger than that in the condensate, thus the 
weak fluctuation in the cavity may strongly influence the dynamics of condensate, which can not be captured by replacing the cavity mode 
by a $c$-number. This mechanism profoundly influences the stability of the fixed points and their corresponding dynamics. We find the cavity 
can play two distinct roles during the dynamics of the condensate. In the oscillating phase, the cavity plays the role of imprinting a gauge potential 
to the effective coupling between the two hyperfine states, giving rise to zero and $\pi$ Josephson oscillation about their fixed points. 
In the trapped phase with small effective coupling, the cavity plays the role of nonreciprocal coupling between the two states. The
stability of the fixed points are tuned from elliptic in a double-well model to stable spiral and unstable spiral due to fluctuation of the
cavity field, as a consequence the condensate can only be trapped in one of the hyperfine state, and the associated dynamics will finally be 
ceased. The transition between these two dynamics can be controlled by the parameters of the cavity field and the driving field.
We estimate the possible observations with experimental accessible parameters. 

\begin{figure}[h]
    \centering
    \includegraphics[width=0.35\textwidth]{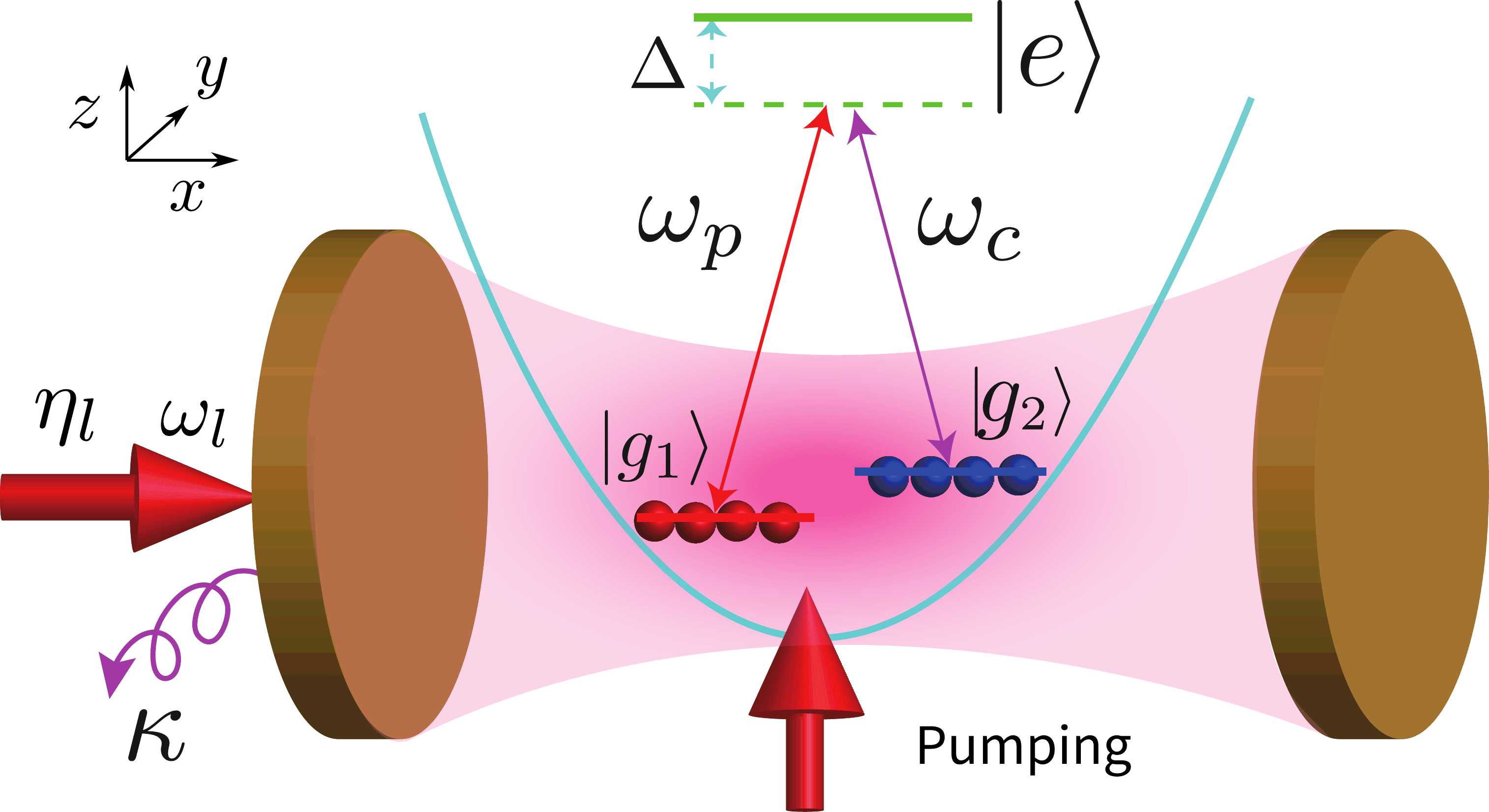}
    \caption{(Color online). Physical model for two-component condensate in a finesse Fabry-P\'{e}rot cavity. The Raman coupling between the two hyperfine states,
        $|g_1\rangle$ and $|g_2\rangle$, is mediated by the pumping field and the cavity mode, far detuned from the excited state $|e\rangle$. $\eta_l$ denotes 
        the driving field, which can be controlled by the driving power $\eta_l=\sqrt{2\kappa P/\hbar\omega_l}$
        \cite{aspelmeyer_cavity_2014,gigan_self-cooling_2006,huang_electromagnetically_2011},
        and $\kappa$ is the decay rate of the cavity.
        For $P=1$ nW and $\kappa=1$ MHz, $\eta_l\approx 8.9\times 10^7$ MHz.}
    \label{fig-fig1}
\end{figure}

We consider a two-component ultracold bosonic atoms trapped in a high finesse cavity (see Fig. \ref{fig-fig1}). 
The ground states $|g_{i}\rangle$ ($i=1, 2$ for two hyperfine states) and excited state $|e\rangle$ are coupled by the pumping laser with coupling strength 
$\Omega_p$ and frequency $\omega_p$, and cavity mode with coupling strength $\Omega_c\cos(k_c x) (\hat a + \hat a^\dagger)$
and frequency $\omega_c$ \cite{hood_atom-cavity_2000}. In the large detuning limit, 
we can safely eliminate the contribution from the excited states via second-order perturbation theory
\cite{maschler_ultracold_2008,brion_adiabatic_2007,dong_photon-induced_2015}, and obtain
\begin{eqnarray}
	H&=& \int \dd {\bf r} \sum_{i=1,2} \left[ \Psi_{i}^\dagger({\bf r})\left( \frac{p^2}{2m}+V_g({\bf r})+H_\mathrm{int}^i +\omega_i \right)\Psi_{i}({\bf r})\right]
    \nonumber\\
    &&+\left[\Omega \cos(k_c x)(\hat{a}+\hat{a}^\dagger)e^{-i\omega_p t}\Psi_{1}\Psi_{2}^\dagger+\Hc \right]
    \nonumber\\
    &&+i\eta_l\left( \hat{a}^\dagger e^{-i\omega_l t}-\hat{a}e^{i\omega_l t} \right)+\omega_c \hat{n}_a,
    \label{eq-hamiltonian}
\end{eqnarray}
where $H_\mathrm{int}^i=\tilde U_0\Psi_{i}^\dagger \Psi_{i}+ \sum_{j\ne i}\frac{\tilde V_0}{2}\Psi_{j}^\dagger \Psi_{j}$
denotes the many-body interaction in the same trap, $\omega_i$ is the energy of each hyperfine state, 
$\Omega=\Omega_c\Omega_p/\Delta$ is the effective coupling from the Raman coupling,
with $\Delta = (\omega_{e} - \omega_1) - \omega_p$ being the energy detuning 
between atoms and pumping laser frequency $\omega_p$, $\hat{n}_a = \hat{a}^\dagger \hat{a}$, with $\hat{a}$ ($\hat a^\dagger$),
is the annihilation (creation) operator of the cavity mode. $\omega_l$ is the driving laser of the cavity. In experiments,
the detuning $\Delta$ is about tens or hundreds of GHz \cite{brennecke_cavity_2008,baumann_dicke_2010}.
The term $\eta_l$ denotes the effective driving strength \cite{barlow_master_2015}.

It is convenient to make a rotation to the above model via a unitary transformation
$\mathcal{U}=e^{-i\omega_l t \hat{a}^\dagger \hat{a}-i\sum_{m=1,2}\omega_mt\op{g_m} }$,
then the effective Hamiltonian may be written as
$\mathcal{H}= \mathcal{U}^\dagger H\mathcal{U} -i\mathcal{U}^\dagger \partial_t \mathcal{U}$.
Under rotating wave approximation, we have
\begin{eqnarray}
	&& \mathcal{H}= \int \dd {\bf r} \sum_{i=1,2} \left[ \Psi_{i}^\dagger({\bf r})\left( \frac{p^2}{2m}
        +V_g({\bf r})+H_\mathrm{int}^i \right)\Psi_{i}({\bf r}) \right] 
    \label{eq-hamiltonian2} \\
	&&+\Omega \cos(k_c x)\left( \hat{a}^\dagger\Psi_{1}\Psi_{2}^\dagger+\Hc \right) 
        + i\eta_l\left( \hat{a}^\dagger -\hat{a}\right)+\delta_c \hat{n}_a,
    \nonumber    
\end{eqnarray}
where the detuning $\delta_c=\omega_c-\omega_l \sim $ kHz (see Fig. \ref{fig-fig1}). In previous models,
without the trapped potential, the coupling between the periodic cavity mode and condensate may excite finite momentum modes in
the condensate, which may lead to bistability between them \cite{zhou_cavity-mediated_2009,zhang_nonlinear_2009,kazemi_controllable_2016,larson_ultracold_2010}. 
Here we consider the dynamics of the condensate in a relative tighter trap, where only the lowest mode needs to be considered. In this case, we 
may write $\Psi_{1} = \psi_b({\bf r}) \hat b$ and $\Psi_{2}= \psi_c({\bf r}) \hat c$ \cite{kittel_quantum_1987}, where $\psi_b$ and $\psi_c$ 
correspond to wave functions of each component. Then,
\begin{eqnarray}
    \mathcal{H}&=& \delta_c \hat{n}_a+ U_0 (\hat{n}_b^2 + \hat{n}_c^2) +V_0\hat{n}_b \hat{n}_c 
    \nonumber\\ &&
    + (g_0 a^\dagger b c^\dagger + \text{h.c.})
    +i\eta_l\left( \hat{a}^\dagger-\hat{a} \right),
    \label{eq-bcdagger}
\end{eqnarray}
where the interaction between the same atoms is assumed to be the same $U_0=\tilde U_0\int \dd {\bf r} |\psi_{b/c}|^4$, and between different
species is denoted as $V_0=\tilde{V}_0\int \dd {\bf r}|\psi_b|^2|\psi_c|^2$. The coupling between the two species is mediated by the second 
term in Eq. \ref{eq-hamiltonian}, that is, $g_0=\Omega\int \dd {\bf r}\cos(k_c x)\psi_b({\bf r})\psi_c^\dagger({\bf r})$. In a harmonic trap, 
let us assume $\psi_b({\bf r}) = \psi_c({\bf r}) \sim \exp(-{\bf r}^2/2\xi^2)$, where oscillating length $\xi = \sqrt{\hbar/(m\bar{\omega})}$.
Then we find $U_0 = {4\pi\hbar^2 a \over m} \int dx \psi_b^4 = 2a\bar{\omega}\sqrt{m\hbar\bar{\omega}}$
and $g_0 =\Omega\exp(-E_R/2\hbar\bar{\omega})$. For parameters in $^{87}$Rb: 
trap frequency $\bar{\omega} \sim 2\pi\times 150$ Hz \cite{dalfovo_theory_1999,ginsberg_coherent_2007,becker_oscillations_2008},
recoil energy $E_R=h \times 3.77$ kHz, Raman coupling
$\Omega=2\pi\times 3.18$ MHz and scattering length $a=20a_0$
($a_0$ is Bohr radius), we estimate $U_0 \sim 24$ Hz, $g_0 \approx 70$ Hz.

We let $g_0$ to be real without losing of generality. The dynamics of the condensate, beyond the steady-state approximation, can be 
captured by the following equations,
\begin{eqnarray}
    \dot {\hat a}&=& i[\hat{a}, \mathcal{H}] = -i\delta_c \hat a-ig_0 \hat c^\dagger \hat b -\kappa \hat{a}/2+\eta_l, \label{eq-4a}
        \\
    \dot{\hat b} &=& i[\hat{b}, \mathcal{H}] =  -2i U_0 \hat{b}^\dagger \hat b^2-iV_0\hat{b} \hat{c}^\dagger \hat{c}-i g_0\hat{c}\hat{a} \label{eq-4b},
        \\
    \dot {\hat c} &=& i[\hat{c}, \mathcal{H}] = -2i U_0 \hat{c}^\dagger \hat c^2-iV_0\hat{c}\hat{b}^\dagger \hat{b}-ig_0 \hat{a}^\dagger \hat{b} \label{eq-4c},
    \end{eqnarray}
where $\kappa$ denotes the decay rate of the cavity mode. The total number of atoms is conserved, that is, $N = n_b + n_c$ and 
$[N, H] = 0$. For a finesse cavity with $Q \sim 10^8 - 10^9$ \cite{thompson_strong_2008, brennecke_cavity_2008, groblacher_demonstration_2009,
burek_high_2014,ritsch_cold_2013}, the decay rate $\kappa \sim $ MHz. We can expand the dynamics of these operators in number-phase representation
\cite{dirac_quantum_1927,paraoanu_josephson_2001}:
$\hat{a}=e^{i\theta_a}\sqrt{n_a^\prime}$, $\hat{b}=e^{i\theta_b}\sqrt{n_b}$ and $\hat{c}=e^{i\theta_c}\sqrt{n_c}$, then the above
equations yield,
\begin{subequations}
    \begin{eqnarray}
        && \dot n_a =  g\sqrt{n_a(1-z^2)}\sin\phi-\kappa n_a+2\eta\sqrt{n_a}\cos(\theta-\phi), \label{eq-5a} \\
        && \dot z=  -2g\sqrt{n_a(1-z^2)}\sin\phi,  \label{eq-5b} \\
        && \dot\theta= -2Uz +V z+2gz\sqrt{\frac{n_a}{1-z^2}}\cos\phi, \label{eq-5c} \\
        && \dot\phi=  \dot\theta +\delta_c+\frac{g}{2}\sqrt{\frac{1-z^2}{n_a}}\cos\phi+\frac{\eta}{\sqrt{n_a}}\sin(\theta-\phi).
        \label{eq-5d}
    \end{eqnarray}
\end{subequations}
Hereafter, we regard $N$ as a normalized factor, and define $g=g_0\sqrt{N}$, $U=U_0N$, $V=V_0N$, $n_a={n_a^\prime}/{N}$, $\eta=\eta_l/\sqrt{N}$. 
Other parameters are $z=(n_b-n_c)/{N}$, $\phi=\theta_b-\theta_c-\theta_a$ and $\theta=\theta_b-\theta_c$. The variables 
$(p, q) = (n_i, \theta_i)$ are canonical conjugate pairs, following standard Hamiltonian equations: $\dot{p}=  \pdv{H_{\rm eff}}{q}$, $\dot{q}=
-\pdv{H_{\rm eff}}{p}$, we have 
\begin{eqnarray}
    \mathcal{H}_{\rm eff}&=& (\delta_c - i\kappa) n_a+g\sqrt{n_a(1-z^2)}\cos\phi + {\mathcal{K} \over 2} z^2
    \nonumber\\
    &&+2\eta\sqrt{n_a}\sin(\theta_a).
    \label{eq-Heff}
\end{eqnarray}
where $\mathcal{K}=U-V/2$. In experiments, while $g \sim $ kHz for number of atoms $N \sim 10^4$,
the last term, $\eta$, is typically of the order of MHz. For the fixed points, one may replace the
operator $\hat{a}$ by a $c$-number, in which case the above model is the same as that in a double
well potential \cite{milburn_quantum_1997,shin_atom_2004,schumm_matter-wave_2005}. For this reason,
we expect coherent oscillation between the two hyperfine states.
However, the interesting point is that the fluctuation of the cavity field, both from the fluctuation
of phase and the number of photons, can have profound consequence to the stability of the 
fixed points. The same picture may be applicable to dynamics in other many-body models in the cavity.

\begin{table}
    \centering
    \begin{tabular}{c|c|c|c|c|c|c}
        \toprule
        &   \multicolumn{3}{c|}{$\mathcal{K}>0$} & \multicolumn{3}{c}{$\mathcal{K}<0$}
         \\ \hline
         FPs &  $\phi^*$ &  $\mathcal{Q} >  4\eta^2g^2$ & $\mathcal{Q} <  4\eta^2g^2$ &  $\phi^*$ &
         $\mathcal{Q} >  4\eta^2g^2$ & $\mathcal{Q} <  4\eta^2g^2$ \\ \hline
         (1)&  0  & Saddle   & Elliptic &  0  & Elliptic & Elliptic  \\ \hline
         (2)&  0  & Saddle  & - &  $\pi$  & Saddle   & - \\ \hline
         (3) & 0  & Stable spiral & - &  $\pi$  & Stable spiral & - \\ \hline
         (4)& $\pi$ & Elliptic & Elliptic &  $\pi$  & Saddle   & Elliptic \\ 
         \hline
    \end{tabular}
    \caption{The stability of the fixed points (FPs). Criteria for these FPs are following. Saddle means $\exists i,j$, with 
		Re$(\lambda_i) < 0$ and Re$(\lambda_j) > 0$. Stable means for all $i$, Re$(\lambda_i) \le 0$. Stable spiral means that 
    	for all $i$, Re$(\lambda_i) < 0$ and Im$(\lambda_i)\ne 0$; while elliptic means $\exists i$ satisfies Re$(\lambda_i) = 0$.}
    \label{tab-tabI}
\end{table}

\begin{figure}[h]
    \centering
    \includegraphics[width=0.45\textwidth]{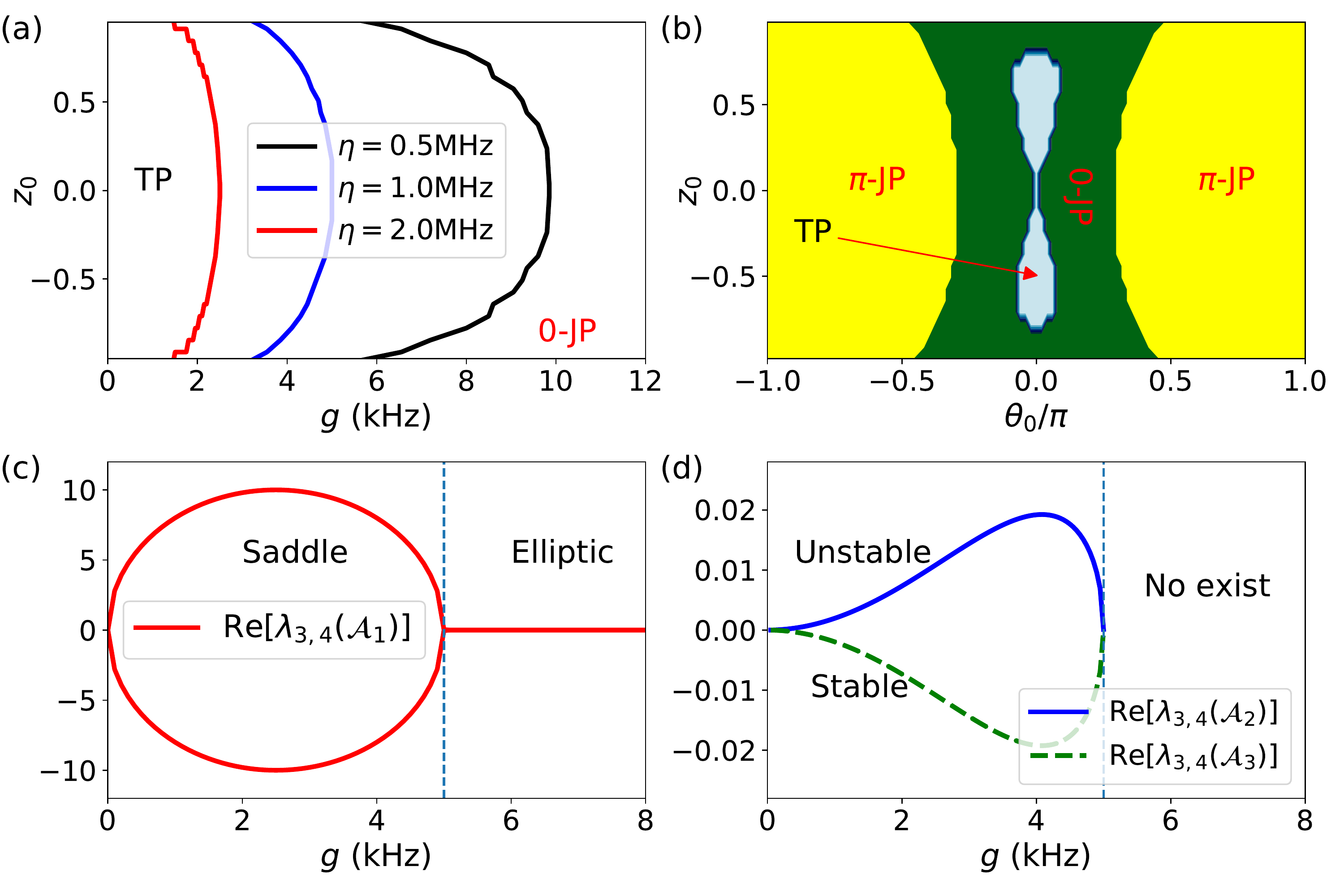}
    \caption{(Color online). Phase diagrams from the long-time dynamics. In (a), we have used initial phase $\theta_0=0$. The condensate
	is at the trapped phase (TP) when $g$ is small; and otherwise in the Josephson oscillation phase (JP). (b) Same as (a),
	but in the $z_0 - \theta_0$ plane with $g = 4.5$ kHz. Besides the Josephson oscillation around $\theta^* \sim 0$, 
	it is also possible to realize the similar phase at $\theta^* \sim \pi$.  
	(c) and (d) The real part of the eigenvalues $\lambda_3$ and $\lambda_4$ as a function of $g$. $\lambda_{1,2} = -\kappa/2$
        for all fixed points, thus is not shown. Parameters are 
        $\eta=1$ MHz, $\kappa=1$ MHz, and $\mathcal{K}=10$ kHz. In experiment with typical $N = 5 \times 10^3$, 
        the parameters $g=5$ kHz, $\mathcal{K}=10$ kHz correspond to $g_0 = 70.7$ Hz, and $U_0-V_0/2 = 2.0$ Hz. }
\label{fig-fig2}
\end{figure}

From Eq. \ref{eq-5b}, the fixed points can only be realized at $\sin(\phi) = 0$. We find this model may host four different superradiant 
fixed points with $n_a^* \ne 0$. To simplify the formula, we let $\delta_c=0$; we will take it back for 
$\delta_c\ll \eta$ in simulation. For $\phi^* = 2n\pi$, $n \in \mathbb{Z}$, with fixed points: (1)
\begin{eqnarray}
    n_a^*= \frac{4\eta^2-g^2}{\kappa^2}, z^*=0, \theta^*=-\mathrm{cos}^{-1} \frac{\sqrt{4\eta^2-g^2}}{2\eta},
    \label{eq-fixxx1}
\end{eqnarray}
(2) for $z^*>0$ and (3) for $z^*<0$ as following
\begin{eqnarray}
    n_a^* = {4\eta^2\mathcal{K}^2 \over \mathcal{Q}},
    z^* = \pm \sqrt{1-\frac{4\eta^2 g^2}{\mathcal{Q}}}, 
    \theta^*=-\mathrm{cos}^{-1} \frac{\kappa \mathcal{K}}{\sqrt{\mathcal Q}},
    \label{eq-fixedchati}
\end{eqnarray}
where $\mathcal{Q} = g^4+\kappa^2\mathcal{K}^2$. For $\phi^*=\pi+2n\pi$, the fixed point is (4)
\begin{eqnarray}
    n_a^*= \frac{4\eta^2-g^2}{\kappa^2}, z^*=0, \theta^*=\mathrm{cos}^{-1} \frac{-\sqrt{4\eta^2-g^2}}{2\eta}.
    \label{eq-fixxx11}
\end{eqnarray}
For $g \ll \eta$, the fixed points (1) and (4) are always exist, while points (2) and (3) may break down when $\mathcal Q> 4\eta^2g^2$. In 
the regime when $\mathcal{K}, g\ll \kappa,\eta$, this inequality is reduced to $g< \kappa \mathcal{K}/2\eta$. 
The normal phase with $n_a^* = 0$ corresponds to the physics without coupling to cavity, thus is trivial to us.
These four fixed points have totally different stability due to fluctuation of the cavity field. If we define
vector ${\bf x} = (n_a - n_a^*, z -z^*, \theta -\theta^*, \phi - \phi^*)$,
then the evolution of ${\bf x}$ after linearization can be written as $\dot{\bf x} = \mathcal{A}_m {\bf x}$, where 
$\mathcal A_m$ ($m=1,2,3,4$) denote the stability matrix of the $m$-th fixed point. Upon a similarity transformation, we find
\begin{eqnarray}
    P^{-1} \mathcal A_m P =
    \mqty(
    -\kappa\sigma_z/4 & u_m\\
    v_m & -\kappa\sigma_z/4
    )-\frac{\kappa}{4}\sigma_0,
    \label{eq-A}
\end{eqnarray}
where $\sigma_{0, z}$ are Pauli matrices. Let $\lambda_j(\mathcal{A}_m)$ being the $j$-th eigenvalue of $\mathcal{A}_m$, then the dynamics of
the condensate are fully determined by the real part of these
eigenvalues. The system is stable only when all real parts are less or equal to zero; otherwise, the system is unstable. A detailed classification
of these behaviors can be found in Refs. \cite{tabor_chaos_1989,shashkin_fixed_1991}. We present the relevant criteria for these fixed points in 
Table \ref{tab-tabI} for sake of convenience.

\begin{figure}[h]
    \centering
    \includegraphics[width=0.45\textwidth]{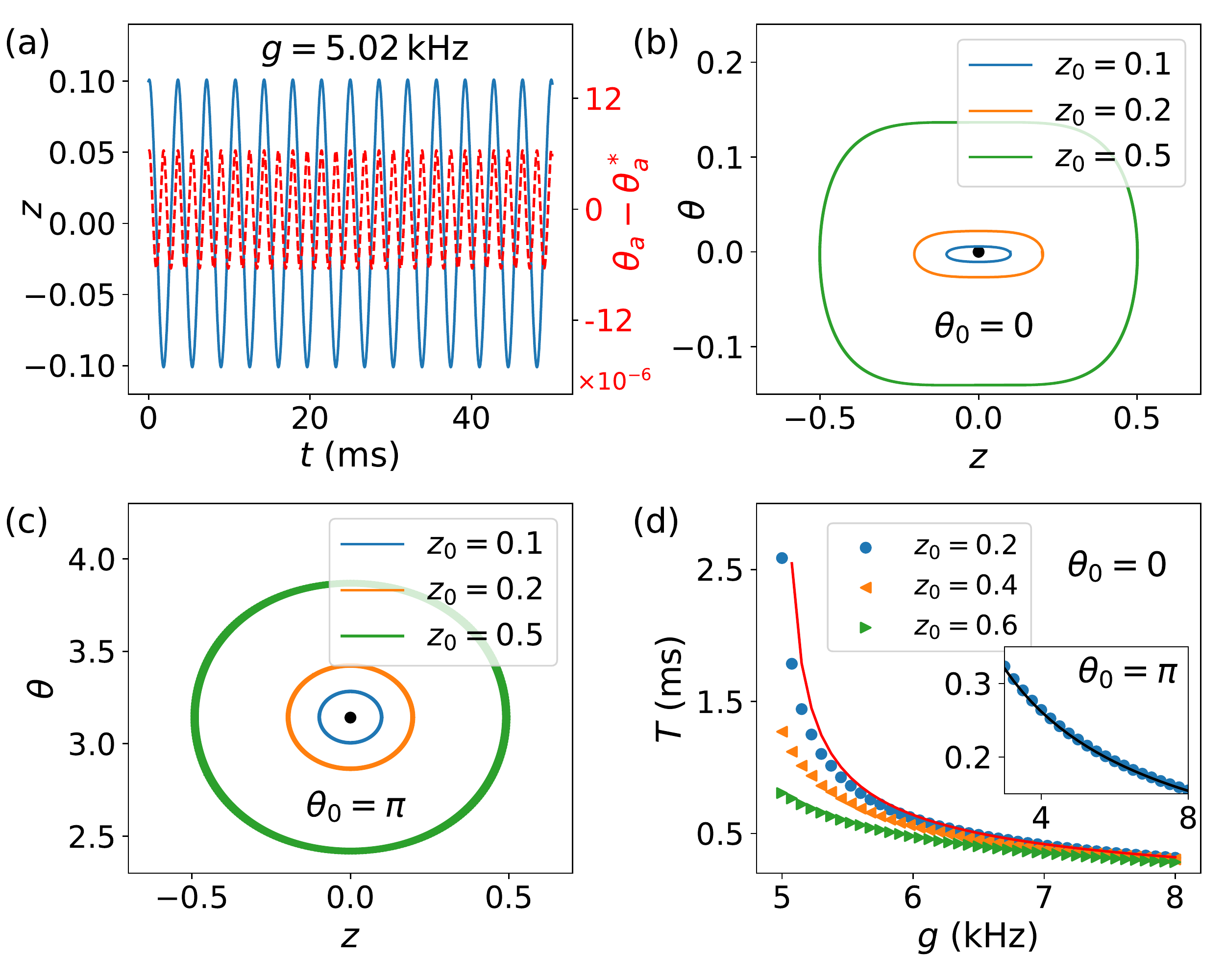}
    \caption{(Color online). Dynamics of the oscillating phase. (a) Periodic oscillation of imbalance $z$ and phase, $(\theta_a-\theta_a^*) N$. 
		(b) and (c) Dynamics in the zero and $\pi$ Josephson oscillating phase, respectively. (d) Oscillating period, $T = 2\pi/\omega_J^{\pm}$ 
		determined by Eq. \ref{eq-omegaJ} (see the solid line). Inset gives the dynamics at $\theta^* \sim \pi$.}
                \label{fig-fig3}
\end{figure}

{\it Josephson oscillating phase}. We plot the phase diagram as a function of initial imbalance $z_0 = z(t=0)$ and effective coupling strength $g$ 
in Fig. \ref{fig-fig2}a. For $\mathcal{K} > 0$, we find two different oscillating phases at (1) $\theta \sim 0$ ($\phi^* = 2n\pi$, termed as zero Josephson
oscillation phase) \cite{albiez_direct_2005,levy_acdc_2007,trujillo-martinez_nonequilibrium_2009}, and at (4) $\theta \sim \pi$ ($\phi^* = (2n+1)\pi$, termed as $\pi$ Josephson phase).
These phases have been searched in superconducting junctions \cite{ryazanov_coupling_2001,kontos_josephson_2002,gingrich_controllable_2016,zhou_silice_2017,bulaevskii_supercon_2017,piano_0_2007}.
Around these two fixed points, $u_m$ and $v_m$ in Eq. \ref{eq-A} read as,
\begin{eqnarray}
    && u_1=\frac{2\eta g}{\kappa}\mqty(0&1\\ -2 &-2), v_1=\left( {4\eta g \over \kappa}-2\mathcal{K} \right)\mqty(0&0\\0 & 1) \label{eq-u1v1}, \\
    && u_4=\frac{2\eta g}{\kappa} \mqty(-2 & -1 \\ 2 & 2), v_4=-({4\eta g \over \kappa}+2\mathcal{K})\mqty(0&0\\0&1). \label{eq-u2v2}
\end{eqnarray}
The eigenvalues are $\lambda_{1,2}(\mathcal A_1, \mathcal A_4)=-\kappa/2$ and
$\lambda_{3,4}(\mathcal A_1) = \pm  \frac{\sqrt{8\eta g(\kappa \mathcal{K}-2\eta g)}}{\kappa}$,
$\lambda_{1,2}(\mathcal A_4)=\pm \frac{\sqrt{8\eta g(\kappa \mathcal{K}+2\eta g)}}{\kappa}i$, indicating of elliptic stability, following criteria 
in Tab. \ref{tab-tabI}. We find the corresponding plasmon frequencies can be written as,
\begin{eqnarray}
    \omega_J^\pm={\sqrt{2J_\mathrm{eff}(2J_\mathrm{eff}\pm 2\mathcal{K} )}},
	\label{eq-omegaJ}
\end{eqnarray}
where the effective coupling $J_\mathrm{eff}= g \sqrt{n_a^*} = 2\eta g/\kappa$. Notice that the dynamics also depends on the initial conditions 
(see Fig. \ref{fig-fig2}a, b). Two typical examples for these two oscillation phases are presented in Fig. 
\ref{fig-fig3}b and c for $\theta^* \sim 0$ and $\theta^* \sim \pi$, respectively. One may understand the stability and dynamical period of
these two dynamics by replacing $\hat{a}$ with a $c$-number from $\dot{a} = 0$ (Eq. \ref{eq-4a}), as was used in literatures
\cite{gati_realization_2006,zhu_geometric_2005,chen_superradiance_2014,chen_quantum_2016}.
Thus the cavity plays the role of imprinting a gauge potential to the dynamics of condensate.  
For $|\delta_c| \ll \eta$, the imprinted phase $\theta_a \approx \frac{2\delta_c}{\kappa}-\frac{g}{2\eta}$, which is always very small, 
but may still observable via homodyne detection \cite{fuwa_experimental_2015,bakker_homodyne_2015,fischer_self-homodyne_2016}.
During this oscillation, the phase fluctuation of the cavity field is estimated to be 
$\delta \theta_a \sim g \delta \phi^2/\eta \sim 10^{-5}$ (see Fig. \ref{fig-fig3}a).  We also find the photon number fluctuation $\delta n_a \sim 
\mathcal{O}(1/N)$.

\begin{figure}[h]
    \centering
    \includegraphics[width=0.49\textwidth]{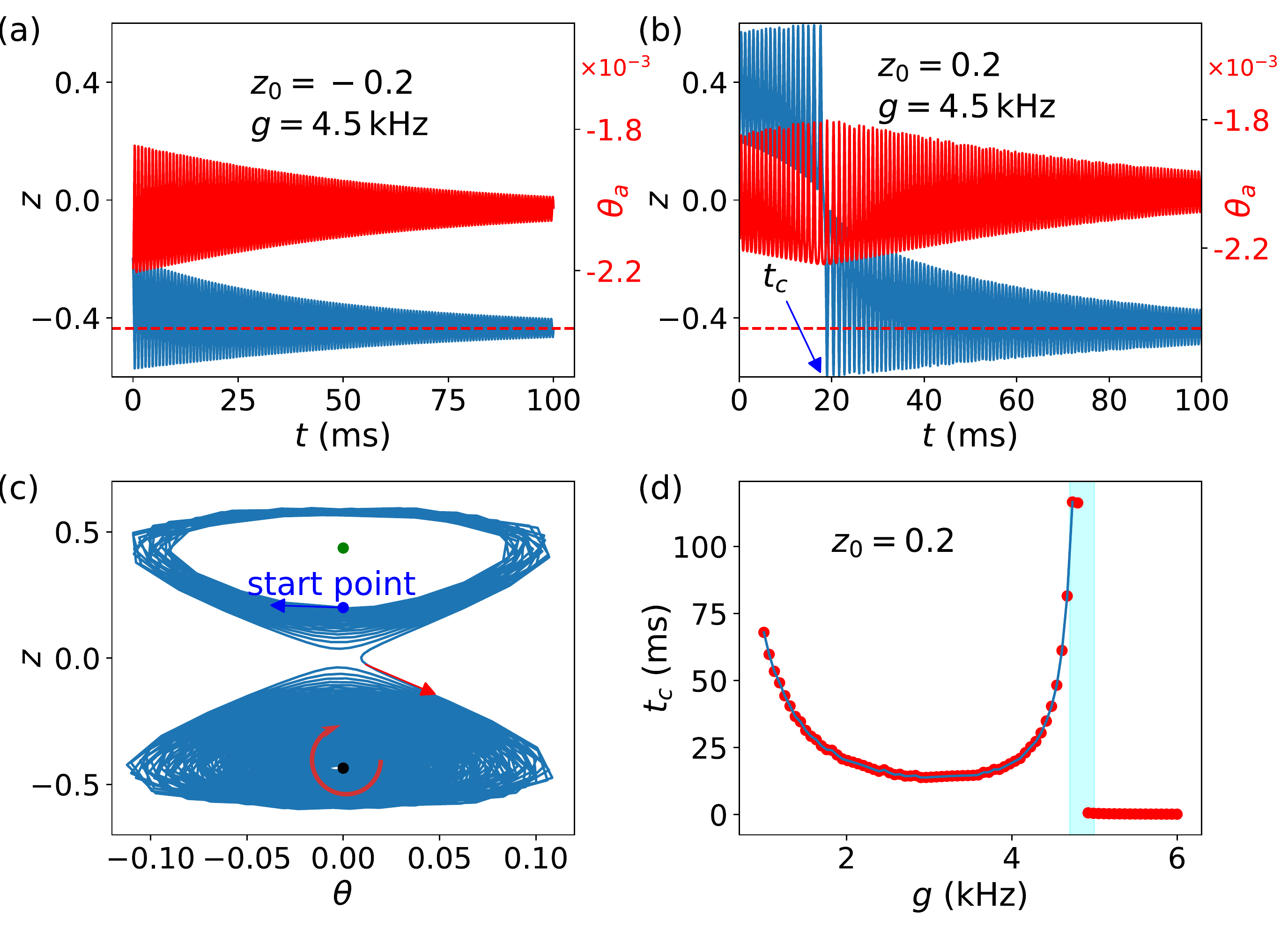}
	\caption{(Color online). Dynamics in the trapping dynamics. (a) and (b) show the dynamics of condensate with initial condition $z_0 < 0$ and $z_0 > 0$, respectively,
        and their corresponding phase fluctuation of the cavity field. In (b), the critical time $t_c$ defines the transition from the unstable spiral point
        to the stable spiral point; see (c) for this transition in phase space. (d) The transition time $t_c$ is typically of the order of tens of ms,
            which will divergent at the phase boundary between trapped and oscillating phases.}
    \label{fig-fig4}
\end{figure}

{\it Trapped phase and dynamics}. In this phase, the cavity plays the role of nonreciprocal coupling between the two hyperfine states. 
The dynamics of condensate will approach a steady state \cite{gati_realization_2006,cataliotti_josephson_2001,javanainen_oscillatory_1986}, in which
the imbalance $z$ and phases will cease to oscillate anymore. Two typical dynamics are presented in Fig. \ref{fig-fig4}a-b. For $z_0 < 0$, the
condensate directly approach the stationary point; however, for $z_0 > 0$, a sudden switch from $z(t) > 0$ to $z(t) < 0$ will happen at
a critical time $t_c$. This feature is totally different from the oscillating dynamics in the self-trapping phase in a double-well potential.
Let $\mathcal{P} =  2\eta^2g^2-\kappa^2\mathcal{K}^2$ and $\mathcal{R}=\mathcal{K}(\mathcal Q-4\eta^2g^2)/2\eta^2g^2$, then in Eq. \ref{eq-A},
\begin{eqnarray}
    u_{2/3}&=& \frac{4\eta^2g^2}{\kappa^2\mathcal{K}}\mqty( 0 & 1 \\ -2 &-2),
    v_{2/3}= 
    \mqty(
    0 & \pm \frac{\mathcal{P}}{8\eta^2 \mathcal{K}} \\
    \mp\frac{\mathcal{P}}{4\eta^2 \mathcal{K}} & \mathcal{R}
    ).
    \label{eq-u23v23}
\end{eqnarray}
The eigenvalues are $\lambda_{3,4}(\mathcal{A}_2,
\mathcal{A}_3) = \pm \frac{2\sqrt{\mathcal{Q}-4\eta^2 g^2}}{\kappa}i+\frac{2\mathcal{P}g^2}{\mathcal{K}^2\kappa^3}\sigma$,
and $\lambda_{1,2}(\mathcal{A}_2, \mathcal{A}_3) = -\kappa/2$, where $\sigma=-1$ and $+1$ correspond to 
fixed point (2) and (3), respectively. These two fixed points exist in the regime when $g<\kappa \mathcal{K}/(2\eta)$, then
$\mathcal P<0$ ensure that $\Re[\lambda_{3,4}(\mathcal{A}_3)]<0$ and $\Re[\lambda_{3,4}(\mathcal{A}_2)]>0$.
This result indicates unstable spiral for $z^* > 0$ and stable spiral for $z^* < 0$, as summarized 
in Tab. \ref{tab-tabI}. The numerical results are presented in Fig. \ref{fig-fig2}d. These two fixed points have opposite real parts (see Fig. 
\ref{fig-fig2}d), due to that the cavity field can induce one of the self-trapping point in the double-well potential model to be unstable 
spiral and the other one to be stable spiral. We find that this change arises from the fluctuation of the cavity field (see Eq. \ref{eq-Heff}).
In Fig. \ref{fig-fig4}c, we plot the trajectory in 
phase space, which first oscillates around the unstable spiral point, and then collapse to the stable spiral point. The transition 
time $t_c$ depends on  $z_0$ and $g$ (see Fig. \ref{fig-fig4}d). Especially, at the boundary between oscillating phase and trapped phase, 
the transition time will divergent. This time scale, of the order of $100$ ms, is accessible in experiments.
In this regime $\theta_a\approx \frac{2\delta_c}{\kappa}-\frac{g^2}{\kappa\mathcal{K}}$.
Since this phase is driven by nonreciprocal 
coupling between the two levels, the similar trapped phase can be found even with finite Zeeman field $\delta_c$.

{\it Conclusion and discussion}. Finally, let us briefly discuss the dynamics in several relevant limits, which always encounter in experiments. 
In the bad cavity limit ($Q \sim 10^6$, $\kappa \sim 100$ MHz), the fast decay of the cavity makes the system favor the trapped phase. However, 
since their boundary is related to $\kappa U$ and $\eta g$, thus their dynamics can be controlled by scatter length, trapped 
potential, as well as parameters of the cavity and driving fields. 
Secondly, as from the effective Hamiltonian $H_\mathrm{eff}$, the fixed points may be interchanged between zero to $\pi$ phases
when $\mathcal{K}$ is changed from positive to negative values. We summarize the properties of the fixed points in this condition in Tab. \ref{tab-tabI}
for a complete comparison. It will change the properties of fixed points (1) and (4), leaving (2) and (3) to be unaffected. Finally, we find that 
during the transition from these two dynamics, the value of of $n_a$ and $\theta_a$ is still a smooth function. Thus in experiments, we have to 
measure the condensate imbalance $z$ using photoemission spectroscopy method \cite{stewart2008using, stewart2010verification, feld2011observation}
to visualize those dynamics.

To conclude, we consider a two-component condensate coupled to a finesse cavity. Since the energy scales in the cavity field is several order of
magnitude bigger than that in the condensate, the small fluctuations in the cavity field may totally influence the stability of the fixed points, 
thus influence the corresponding coherent dynamics. In our model, we find two distinct roles played by the cavity field. In one hand, it may imprint
a gauge potential to the condensate, and on the other hand, it plays the role of nonreciprocal coupling between the two hyperfine states, which tunes the
elliptic fixed points to unstable and stable spiral fixed points. These results demonstrate a novel way to engineer the dynamics by tuning the 
bistability of the fixed points. 

\textit{Acknowledgements.} We thank Chang-kang Hu, Xiang-fa Zhou and Han Pu for valuable discussion. This work is supported by the National Youth Thousand Talents
Program (No. KJ2030000001), the USTC start-up funding (No. KY2030000053), the NSFC (No. GG2470000101).

\bibliography{ref}

\end{document}